**Space-Time Covid-19 Bayesian SIR modeling in South Carolina**

Andrew B Lawson

Joanne Kim

Department of Public Health Sciences, Medical University of South Carolina, Charleston, SC, USA

**Abstract**

The Covid-19 pandemic has spread across the world since the beginning of 2020. Many regions have experienced its effects. The state of South Carolina in the USA has seen cases since early March 2020 and a primary peak in early April 2020. A lockdown was imposed on April 6[th] but lifting of restrictions started on April 24[th]. The daily case and death data as reported by NCHS (deaths) via the New York Times GitHUB repository have been analyzed and approaches to modeling of the data are presented. Prediction is also considered and the role of asymptomatic transmission is assessed as a latent unobserved effect. Two different time periods are examined and one step prediction is provided.

1) **Introduction**

The Covid 19 pandemic has spread across the world since the beginning of 2020. In the US there has been a large increase of cases (symptomatic) and deaths attributable to Covid19. Many serious issues remain about predictive modeling, interventions and vaccines (for a recent overview see Banks et al, 2020). Various data sources are now available that provide daily counts of cases and fatalities for this disease. In this report we demonstrate the use of a Bayesian susceptible-Infect-removed model with spatial geo-referencing at the county level in South Carolina. The model specifies a transmission rate within each county as well as optionally including neighboring county transmissions. Asymptomatic cases are assumed to be a proportion of symptomatic and are included in the transmission model. An accounting equation is assumed where the susceptible population is decremented by infectives and removals. Predictions based on this model are possible, but the focus of this paper is mainly on the model development and fitting. As the pandemic of Covid-19 develops, a variety of approaches to modeling its spread are being presented ( e. g. Calavetti et al, 2020; Calafiore, 2020; Rouabah et a;, 2020). Many of these models are time series-based, and do not incorporate spatial structure in the modeling process. Within Bayesian modeling, few published examples exist where the asymptomatic process is modelled (Zhou et al , 2020). However it is clear that cross boundary transmission must occur (at least at the county level) and so ignoring this must limit the ability of the models to account for this feature. In addition the use of random effects to allow for confounding is often ignored. In the present approach we examine a range of models with different assumptions about transmission and spatial dependence both in terms of fixed predictors and

2) **Data sources**

Both Johns Hopkins University CSSE (https://coronavirus.jhu.edu/map.html) and New York Times (NYT) (https://github.com/nytimes/covid-19-data ) provide data resources at the county level within states of the US. In this report we have analysed the raw data hosted by NYT provided from NCHS (deaths) and state health departments (cases). The data in general is under-ascertained and potentially recorded at time of analysis with error due to misreporting and changes in recording. This is especially true for death data where originally only hospital deaths from Covid19 were recorded and reporting lag time is also present.

It is known that Covid19 has associated asymptomatic cases as well as symptomatic. Asymptomatic cases are not reported usually, could be pre-symptomatic or not, and so cannot be observed unless widespread repeated surveys are carried out in the population. Hence, as these cases can infect others, they must be accounted for in any model. Deaths from Covid-19 for asymptomatic cases are not recorded at all, as they are not tested.

In what follows we assume that the case count (symptomatic) and death count at county level for each day, are the data to be modelled. Of course the under-ascertainment of both case count and deaths is a caveat in the subsequent interpretation. Handling of asymptomatic and under-ascertainment is discussed in the modelling section.

### 3) Model Development

Susceptible–Infected-removed (SIR) compartment models have been used extensively for infection modeling (Keeling and Rohani, 2008). Here the seasonal influenza SIR models by Lawson and Song (2010) (L&S) which were applied spatially to county level biweekly count data (C+ve laboratory notifications) for a single flu season, have been modified and employed to model the Covid-19 count data. These were developed for a spatial application from the time series models of Morton and Finkenstädt (2004) (M&F) which were applied to biweekly measles data for English and Welsh towns for the period 1944 – 1963.

Our data is daily symptomatic ($y_{ij}^{sym}$) case counts and daily death ($d_{ij}$) counts. Asymptomatic ($y_{ij}^{asym}$) case counts are considered as a proportion of $y_{ij}^{sym}$ counts. Denote the spatial and temporal study units as $i = 1,..,m$ counties; $j = 1,....,T$ days.

We also assume that the county susceptible population ($S_{ij}$) at any given time is available. As this does not vary much over a period of a few months then we can assume that the susceptible population at the beginning of the monitoring can be used as a baseline. We do not have access to disease data on sub groups within the population, such as age or gender, and so we must assume a common population base.

The susceptible population in each spatial region (county) is quite large compared to the case count and so it is reasonable to assume a Poisson model for the symptomatic counts, i.e.

$$y_{ij}^{sym} \sim Pois(\mu_{ij})$$
$$\mu_{ij} = S_{ij} f(y_{i,j-1}^{sym}.......) \qquad (1)$$

Note that the conditional independence in the Bayesian hierarchical formulation allows us to assume independence and hence a Poisson data model. An alternative to use a binomial spatial model would be possible also. This doesn't prohibit any extra variation or overdispersion which often arises, as in the Bayesian formulation such variation can be accommodated at higher levels of the hierarchy via prior distributions.

The Poisson mean (1) ( as for M&F) is assumed to be a function of current susceptibles and some function of previous infectives as well as other propagating effects (e. g. random effects or socio-economic predictors/covariates). Note that we also assume that the observed symptomatic cases are to be modeled as opposed to the unobserved total case count. M&F assumed that you can use a binomial model with unobserved total count as denominator (such as $y_{ij}^{sym} \sim bin(\lambda, y_{ij}^{tr})$ where $\lambda$ is a scaling factor) and the total (latent) count is subsequently modelled. However this is a computationally very expensive undertaking, as none of the $y_{ij}^{tr}$ are observed and the complete latent field would have to be estimated. It is simpler to model the observed count and use a scaling factor to estimate the true count at a later stage. This was the approach adopted by L&S and is commonly assumed by health departments during conventional influenza seasons.

In addition to the data model for new counts we also assume an accounting equation which is used to update the susceptible population at each time point. We assume a susceptible – Infected -removed (SIR) model, which is the simplest compartment model. SEIR models including an exposed category are infeasible due to the lack of information about exposure transitions. We assume that susceptibles, if infected, move to the infected category (symptomatic and asymptomatic cases) and either are subsequently removed via death or other forms such as recovery. In the Covid19 data available currently for the US we do not have information on recovered individuals nor direct evidence of asymptomatic cases. As these can affect the progression of the infection we must make assumptions about them. We assume in what follows that the recovery rate is a proportion of the infective count.

The accounting equation is:

$$S_{ij} = S_{i,j-1} - y_{i,j-1}^{tr} - R_{i,j-1}$$

where the current susceptible population is updated by decrementing the previous population by (true) infected count and removal. In our modeling we replace the $y_{i,j-1}^{tr}$ by the sum of asymptomatic and symptomatic cases for that period ( $y_{i,j-1}^{sym} + y_{i,j-1}^{asym}$ where $y_{i,j-1}^{asym} = \varphi y_{i,j-1}^{sym}$ where $\varphi$ is the proportion of asymptomatics in the population. ). This is equivalent to the extended latent influx process of M&F. Also removal is defined as $R_{i,j-1} = Rc_{i,j-1} + d_{i,j-1}$ where $Rc_{i,j-1} = \beta_{rc} y_{i,j-1}^{sym}$ is the recovery rate as a function of symptomatic cases. The deaths ($d_{i,j-1}$) are observed. We don't include asymptomatic recovery, as this is uncertain and unmeasured, although this could contribute.

### 3.1 Models for $f(...)$

In equation (1) above, the $f(...)$ must be specified. We have assumed a variety of models for the dependence on previous cases. Define $Ty_{i,j-1} = y_{i,j-1}^{sym} + y_{i,j-1}^{asym}$ and

*Model* 1) $\log f(Ty_{i,j-1}) = b_0 + b_1 \log[Ty_{i,j-1}] + b_i$

*Model* 2) $\log f(Ty_{i,j-1}) = b_0 + b_1 \log[Ty_{i,j-1} + \sum_{l \in \delta_i} Ty_{l,j-1}] + b_i$

*Model* 3) $\log f(Ty_{i,j-1}) = b_0 + b_1 \log Ty_{i,j-1} + b_2[x_i] + b_i$

*Model* 4) $\log f(Ty_{i,j-1}) = b_{0j} + b_1 \log Ty_{i,j-1} + b_2[x_i] + b_i$

*Model* 5) $\log f(Ty_{i,j-1}) = b_{0i} + b_1 \log Ty_{i,j-1} + b_2[x_i] + v_i$

where $\sum_{l \in \delta_i} Ty_{l,j-1}$ is the sum of *neighboring* infectives *and* $x_i$ is a poverty covariate (% *of population under the poverty line*) and $b_i$ is a spatial random effect.

Note that $b_0$ is a transmission rate (on the log scale) and we could estimate $R_0$ based on $\exp(b_0)/\bar{R}$ but when variants are used such as time dependent $b_{0j}$ or spatially dependent $b_{0i}$ then this is inappropriate. Model 1 has simple prior dependence on the last infective count in the region, and a spatially structured effect $b_i$. Model 2 has an added neighborhood dependence, which could represent cross-boundary infection. Model 3 is as for model 1 but with a county level poverty covariate, and model 4 and 5 include that covariate but also have different transmission parameterisations. Model 5 drops the spatial random effect as the transmission is spatially dependent, but includes a uncorrelated spatial effect ($v_i$). The inclusion of the spatial effects is really to allow for residual confounding whether spatially structured or not.

### 3.2 Prior distributions

The following prior distributions were assumed for the free model parameters. For fixed $b_0$ and $b_1$ and $b_2$ we assume a zero mean Gaussian prior distribution with precision having a gamma (2,0.5). The spatially structured effect is assumed to have a markov random field prior distribution (ICAR) whereby

$$b_i |_{j \neq i} \sim N(M_i, \tau_b^{-1}/n_{\delta_i})$$
$$M_i = \sum_{l \in \delta_i} b_l / n_{\delta_i}$$

The precision has a gamma prior distribution also. The ICAR model was also assumed for the $b_{0i}$ in model 5. The temporally dependent $b_{0j}$ was allowed to have an uncorrelated zero mean Gaussian prior distribution with precision which also had a gamma (2,0.5) distribution. Prior distribution choice is as to be as non-informative as possible, where appropriate. The gamma prior distributions on precisions are weakly informative (STAN wiki 2020).

### 3.3 Computation

Our Bayesian hierarchical models and variants can be conveniently implemented via the Bayesian modeling package `nimble.` This package is available on R and allows the modeling of relatively complex hierarchical models with spatial signatures. The package includes both ICAR and PCAR variants and is based on parsed versions of BUGS code but composes its own C++ samplers. Its benchmark speed is exceptional for a range of problems.

All models were run to convergence, and checked using single chain Geweke diagnostics.

### 3.4 Model Selection

The deviance (D) based measure DIC (Spiegelhalter et al, 2014) was used, with the effective number of parameters computed from the conservative $\hat{p}D = \text{var}(D)/2$, which is always available from MCMC samples. Lower values of DIC suggest improved fit.

## 4.0 Results

### 4.1 Daily Data: January 22$^{nd}$ to April 12$^{th}$

The first analysis undertaken is for data from the period January 22$^{nd}$ to April 12$^{th}$. This consists of 82 days. Most of the early incidence is sparse and so many days through January and February have zero counts for virtually all counties. Figure1 displays the county map of South Carolina.

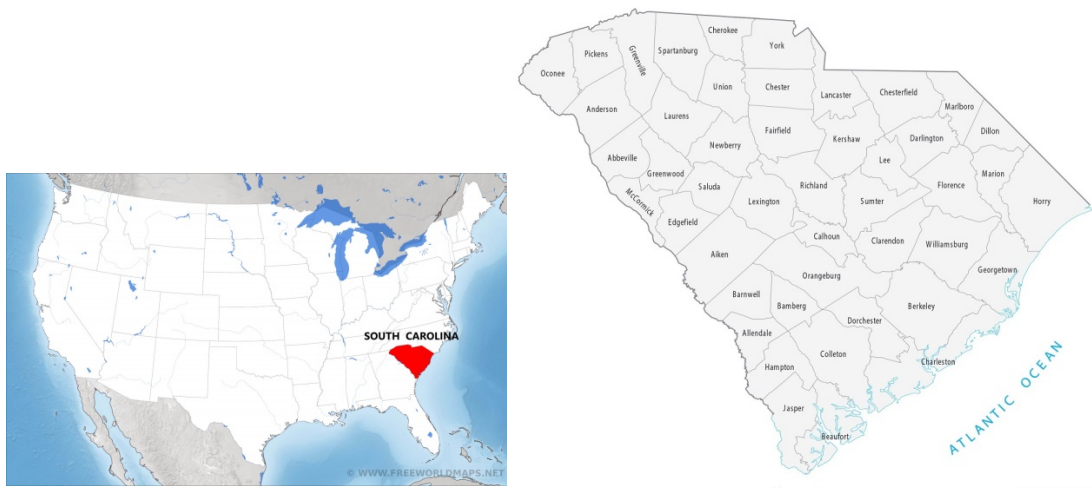

Figure 1 South Carolina: (a) location map in USA, (b) county map

The main urban areas in the state are in Greenville, Spartanburg, Richland, Charleston and Horry counties. As early as March 7$^{th}$ cases appeared in Kershaw county, which is close to the state capital county of Richland. After this, most counties experience incidence which quickly rises to various peaks in early April. Figure 2 displays the symptomatic count and death time profiles for four counties.

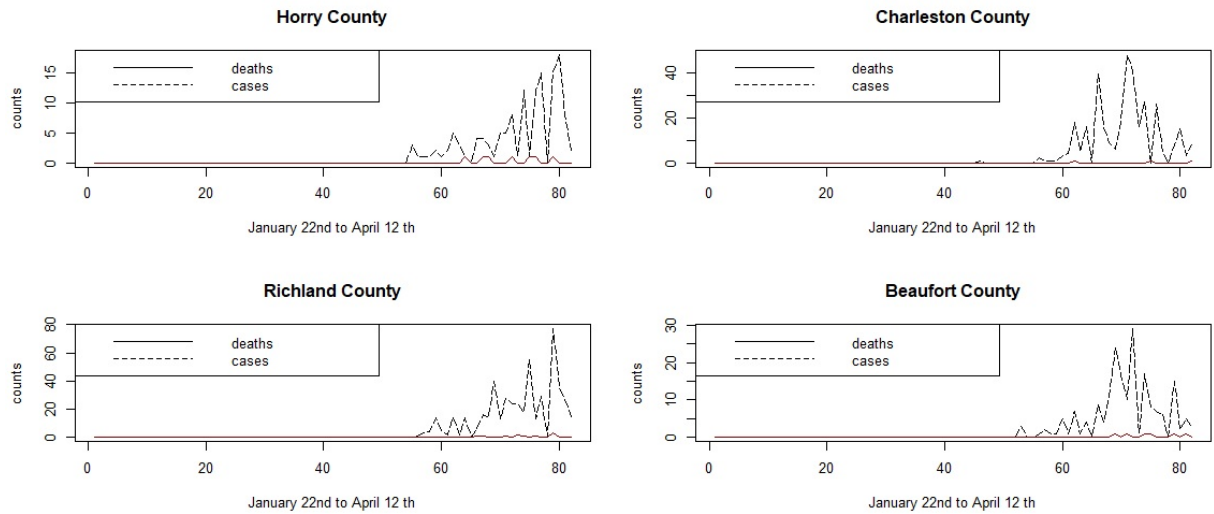

Figure 2 Daily counts of Covid-19 positive symptomatic cases and deaths for four SC counties for the period January 22$^{nd}$ to April 12$^{th}$.

It is notable that the coastal communities of Horry and Beaufort counties, which have an older demographic, have earlier initiation of infection and earlier deaths than the main urban areas of Charleston and Richland counties.

The comparison of the models 1) – 5) yielded the following DICs in Table 1.

Table 1

| Model | DIC | pD | Comment |
|---|---|---|---|
| 1 | 19449.2 | 72.16 | |
| 2 | 10,321,163.0 | 2368.1 | |
| 3 | 19431.0 | 55.99 | |
| 4 | 14448.3 | 130.0 | Asym rate =0.25 |
| 4b | 14428.8 | 128.2 | Asym rate =0.5 |
| 4c | 14504.5 | 125.1 | Asym rate = 0.1 |
| 4d | 15057.6 | 36.97 | Model 4 minus ICAR |
| 5 | 19454.9 | 75.34 | |

Note : 4c is as for model 4, but without the spatially structured ICAR component.

It is clearly the case that the lowest DIC is for model 4 and that models 1, 2, 3, 5 are not competitive. Model 4 with its time dependent transmission rate is clearly better in describing these data than other models. Model 4 variants were examined also. Model 4b allowed the asymptomatic rate ($\varphi$) to change from 0.25 to 0.5 in the model. While model 4c assumes a rate of 10%. The lowest DIC is for model with 50% asymptomatic, although the 25% rate has only a slightly higher DIC. Model 4d variant simply removed the ICAR spatial component and also yielded a larger DIC. It's also notable that the ICAR component seems to be important in the description of the residual confounding in these data. The Appendix displays the nimble code for the model 4).

Figure 3 displays the posterior mean estimates of the mean infection risk profiles for Richland and Kershaw counties. Figure 4 displays the posterior mean risk profiles for Beaufort and Charleston counties. It is clear that the SIR model with time dependent transmission rate and spatial confounding appears to fit the observed infection profiles reasonably well.

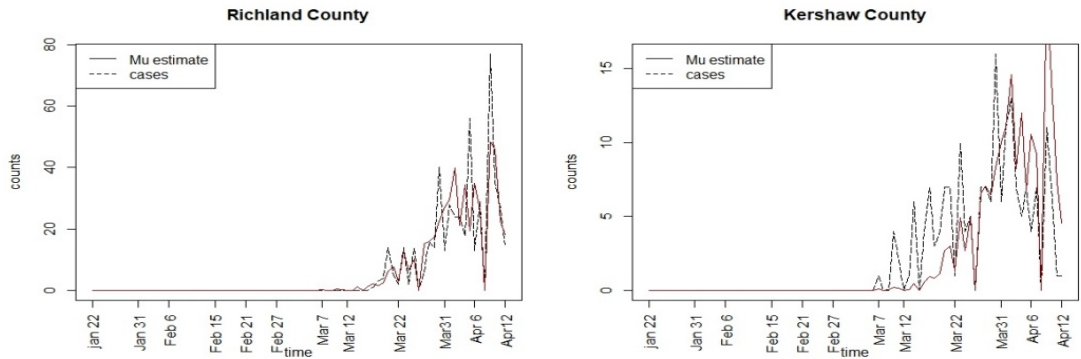

Figure 3 Posterior mean infection risk profiles for Richland and Kershaw counties for model 4

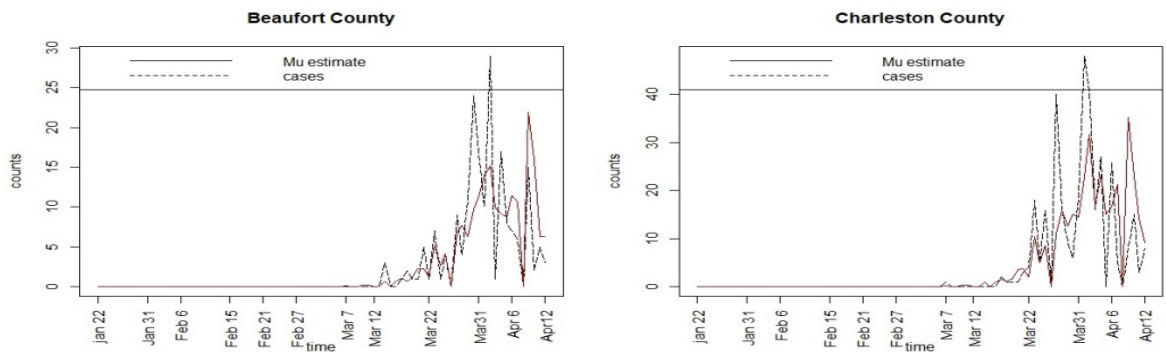

Figure 4 Posterior mean infection risk profiles for Beaufort and Charleston counties for model 4.

The overall DIC for the model is 14428.8, with pD of 128.2 which is considerably lower that alternate models considered.

### 4.1 Daily Data: April 2$^{nd}$ to June 29$^{th}$

The early period data demonstrates the beginnings of the outbreak where there were zero cases until March and then a large increase during March and April. Crucially, the state was not completely locked down until April 6$^{th}$ when a stay-at-home order was issued. By this time there had been a large increase in cases. Partial lifting occurred on April 24$^{th}$ when some businesses were allowed to open. Restaurants opened on May 1$^{st}$. Following this date a gradual rise in cases numbers occurred as social distancing and face covering recommendations were apparently not adhered to, both in the general population and in hospitality industry venues.

A range of models as per the earlier time period were fitted to these data. Table 2 displays the goodness of fit results. For these data model 3 provides the lowest DIC, although model 1 and model 5 have close values. Model 3 assumes an asymptomatic rate of 25% and includes the poverty covariate. Variants of model 3 with 50% asymptomatic and also no ICAR component had much higher DICs. The posterior mean parameter estimates (and SDs) for the different components are $b_0$: 8.036 (0.064), $b_1$: -0.542 (0.0052), $b_2$: -0.800 (0.0031). These are well estimated and suggests an average transmission of 0.6 over the whole period, and a dependence on % poverty of 0.449. It is notable that the asymptomatic rate assumption of 25% proved to yield a better fit than 50%.

Table 2

| Model | DIC | pD | Comment |
|---|---|---|---|
| 1 | 41197.5 | 70.8 | |
| 2 | 10359807 | 533.3 | |
| 3 | 41193.3 | 67.01 | Asym rate=0.25 |
| 3b | 43872.3 | 63.4 | Asym rate =0.5 |
| 3c | 43319.84 | 2.95 | Model 3 minus ICAR effect |
| 4 | 45475.3 | 165.7 | Time varying |
| 5 | 41195.1 | 69.0 | Space varying |

Figures 5 and 6 display the case count profiles and posterior mean estimates for a selection of counties around the state, for model 3. It is notable that the estimates track the case counts in a smooth manner but tend to adjust upward the lower counts during April and May but are downward biased during June, when the counts increases markedly. Figure 7 displays the estimated mean square error profile for the fitted model 3, computed at each time point. This displays the variation in fit and suggests a good fit in early days but also supports that the model fits progressively less well when counts spike.

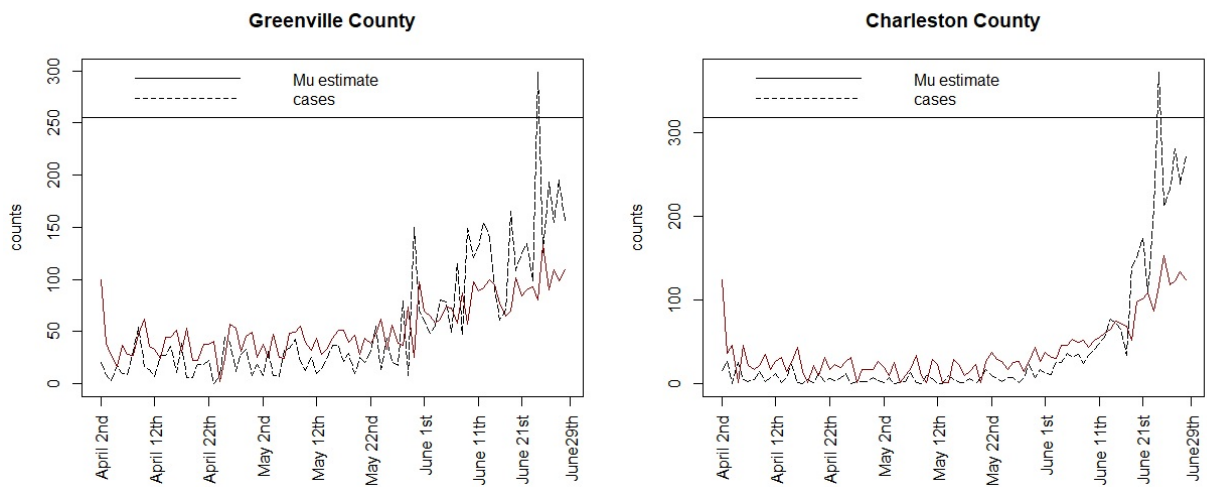

Figure 5 Greenville and Charleston county posterior mean estimates and case counts for model 3

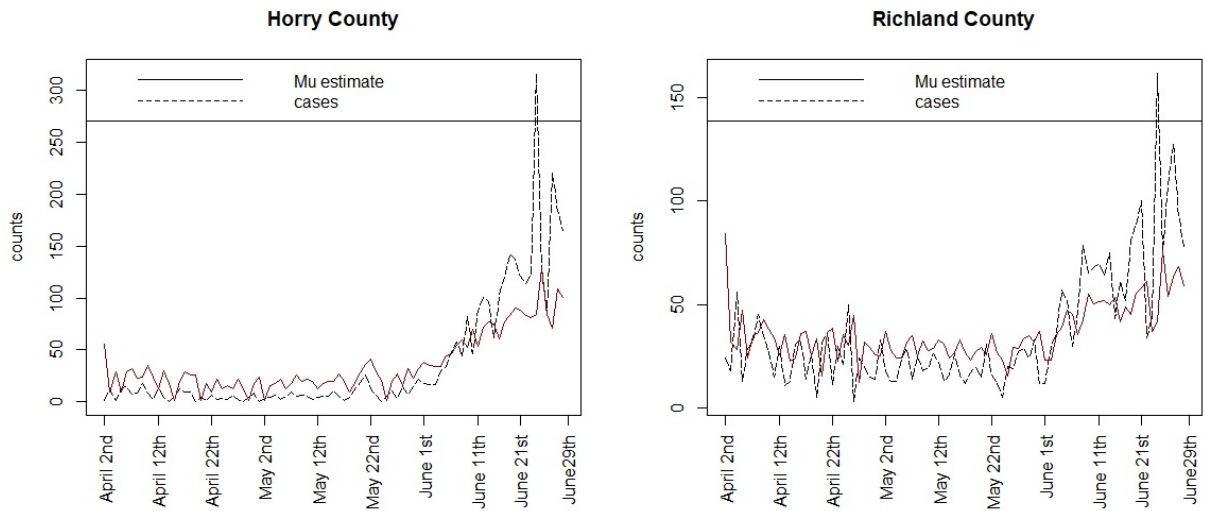

Figure 6 Posterior mean estimates and case counts for Horry and Richland counties for model 3.

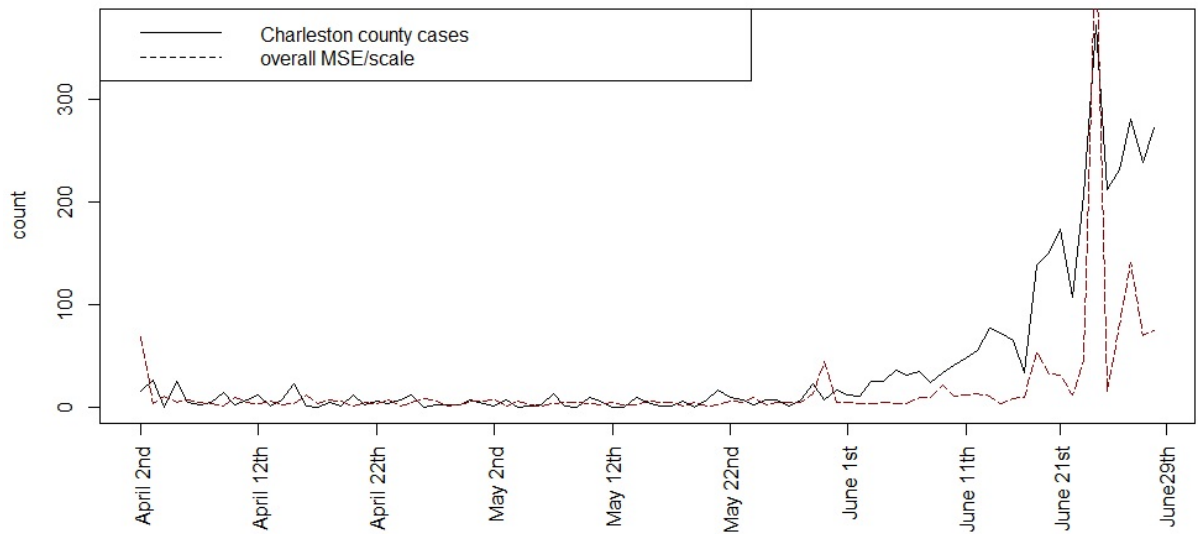

Figure 7 Mean square error (MSE) over time for the fitted model 3 estimates for Charleston county

## Prediction

Both within sample and beyond sample prediction could be important in the context of infection modeling. The mean square predictive error is a measure of loss between predicted model and observed outcomes within sample (Gelfand and Ghosh, 1998). Figure 8 displays the MSPE for the fitted model 3 for these data, computed at each time point. It is noticeable that the error is reasonably small until a large peak or jump in risk occurs although the error does increase with increasing spiking of risk.

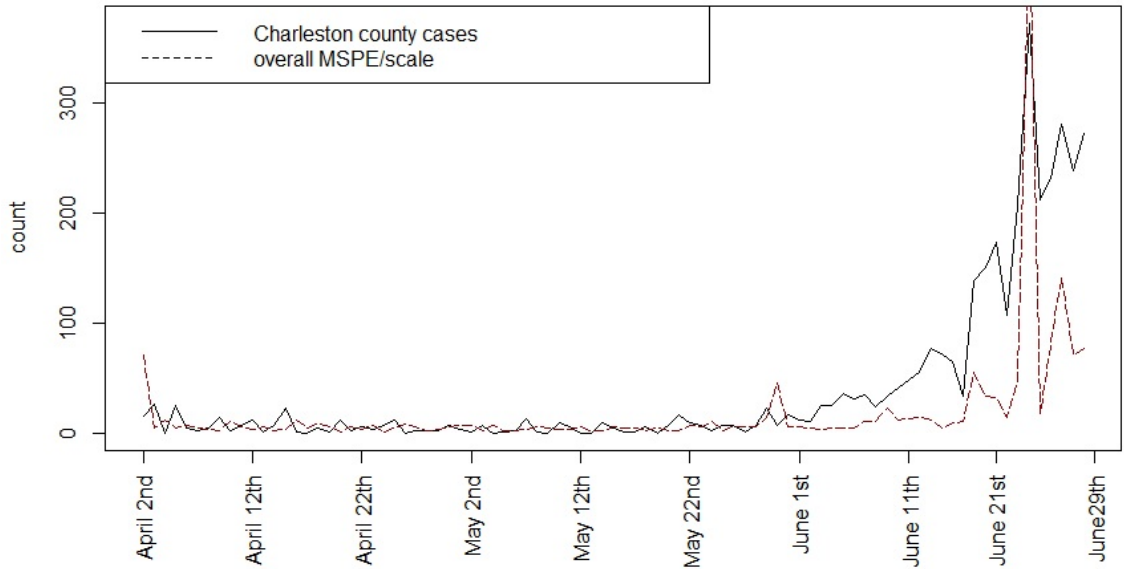

Figure 8 Mean square predictive error (MSPE) over time for fitted model 3 for Charleston county

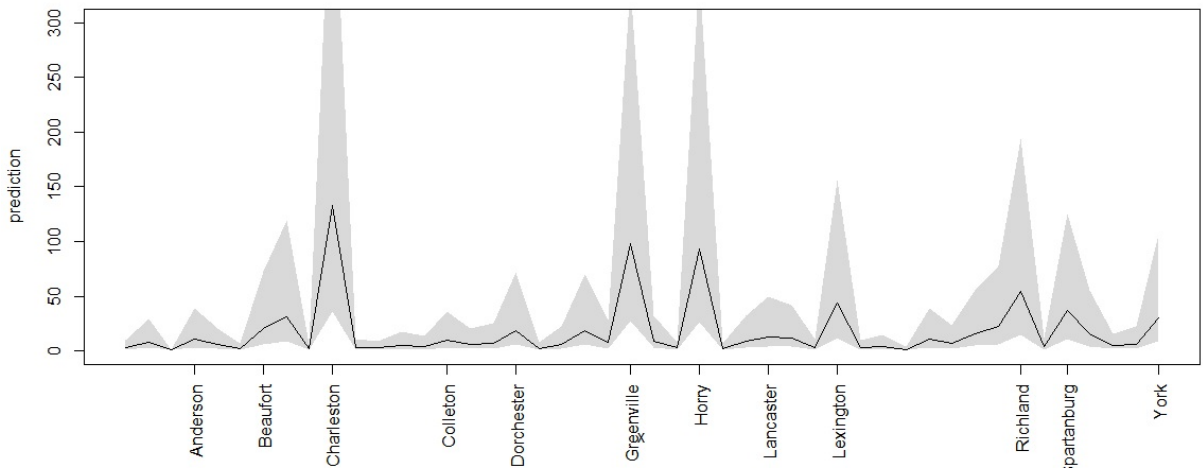

Figure 9 One step predictions with 95% Bayesian credible intervals for SC counties for June 30th

As part of the predictive capability of the Bayesian space-time models it is possible to consider 'step ahead' forecasts for these data. There are a number of ways that predictive inference can be used for temporal prediction. Here a simple one step (one day) prediction is made using the final day (June 29th) to provide accounting update and model dependence on the final observed counts. Figure 9 displays the prediction of SC counties for June 30th, with associated 95% credible interval (shaded).areas. In particular Charleston, Greenville, Horry, Lexington, Richland and Spartanburg have wide ranges. Its clear that the prediction interval is highly asymmetric with a long tail in the higher risk levels.

Any prediction beyond one step that would involve extra variation due to the use of estimated quantities in the modeling process and hence uncertainty would be propagated.

## 4.2 Models fitted to smoothed 3 day averages

Smoothed 3–day average data provides for some removal of anomalies in the count data. Anomalies such as weekend reporting delays and changes in recording, could be ameliorated to some degree by such smoothing. The smoothed counts could be assumed to be Poisson distributed (even if averaged) and the smoothing will induce temporal correlation that can be accommodated by the models. However, for consistency, the 3 day average of the centered counts was computed, which produced smooth estimates at each daily time point. This should also reduce the differentiation of neighboring areas and anomalies relates to weekend non-reporting.

Figure 10 displays the smoothed case data and death count for 4 SC counties for the first period January 22$^{nd}$ to April 12$^{th}$. Figure 11 displays the smoothed case and death data for April 2$^{nd}$ – to June 29$^{th}$.

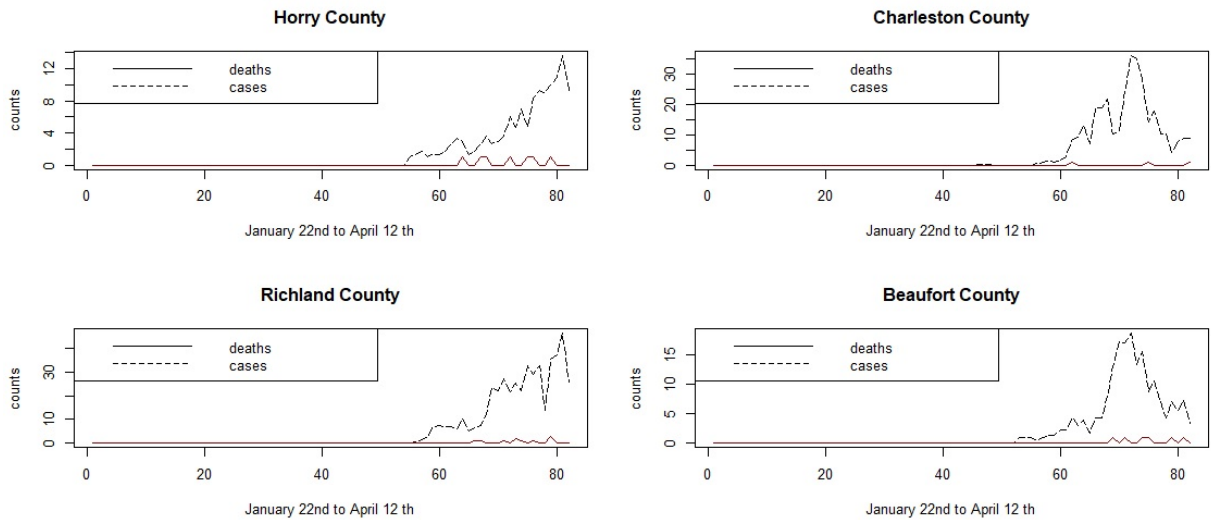

Figure 10 Time plots of smoothed case averages and deaths for the period of January 22nd to April 12th

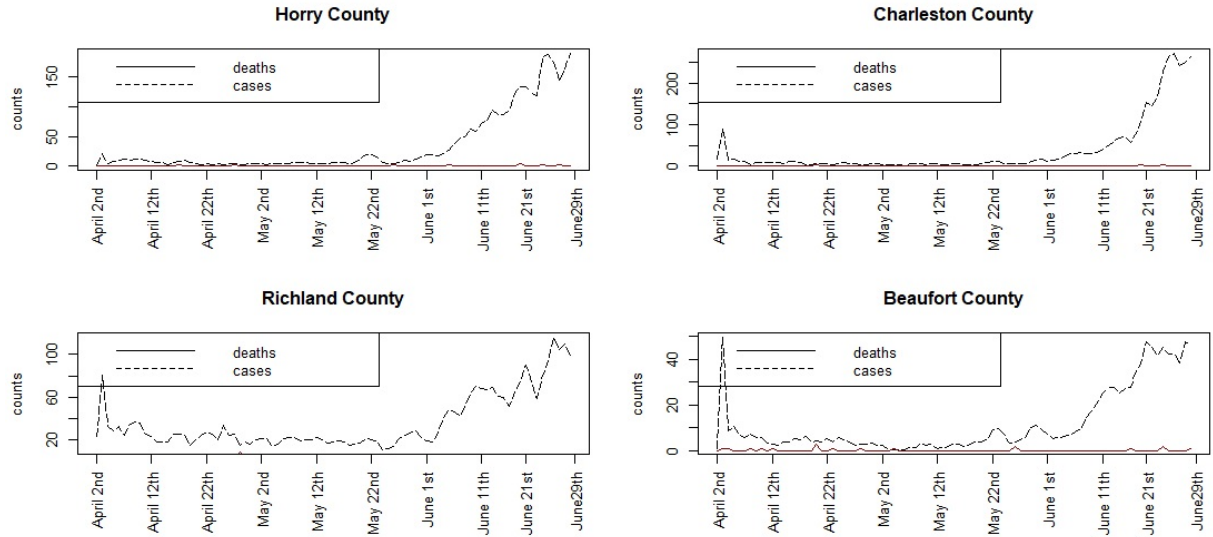

Figure 11 Time plots of smoothed case averages and deaths for the period April 2nd to June 29th

Assume that $y^{sm}_{ij}$ is the 3 day smoothed symptomatic case average. As this is now continuous and strictly positive, a different data model must be assumed. A log normal model for the 3 day average was adopted so that $y^{sm}_{ij} = \exp(y^*_{ij})$; $y^*_{ij} \sim N(\mu_{ij}, \tau_y^{-1})$. A range of models for $\mu_{ij}$ have been examined as in the non-smoothed case.

Both the original January-April and April-June data have been analyzed using the models 1) - 5) The results are found in Table 3.

Table 3

| Model | Jan 22nd - April 12th 2020 | DIC | pD | April 2nd - June 29th 2020 | DIC | pD |
|---|---|---|---|---|---|---|
| 1 | | 99891 | 804.5 | | 132103 | 1200.6 |
| 2 | | 99939 | 875.4 | | 132168 | 1260.6 |
| 3 | | 99858 | 767.5 | * | 132093 | 1204.6 |
| 4 | | 100127 | 1060.0 | | 132953 | 1941.0 |
| 5 | * | 99785 | 729.4 | | 132350 | 1415.9 |

It is notable that for the earlier data the model with the spatially varying intercept parameter (model 5) has lowest DIC. This suggests that spatial structure plays an extra role in the earlier period. In contrast in the later period the model with only the poverty covariate and spatial CAR effect (model 3) has lowest DIC. For the later period a one step ahead forecast has also been obtained from the smoothed model (3).

Figure 12 displays the smoothed 3 day prediction for the later data set. It is clear that Charleston, Greenville, Horry and Richland have the largest intervals now and largest predicted counts, and the influence of smoothing has reduced the prediction for Lexington and Spartanburg.

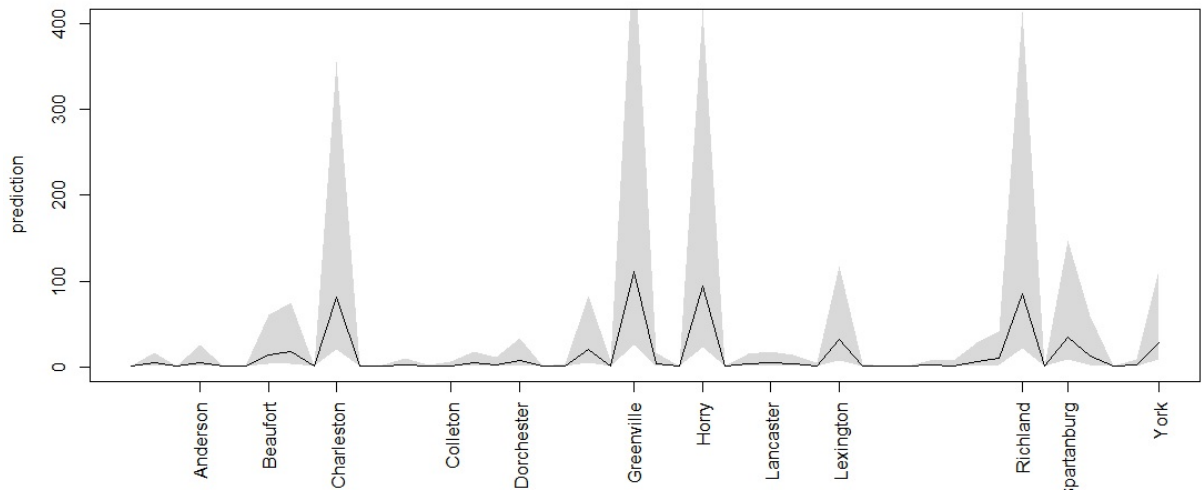

Figure 12 June 30th prediction for smoothed 3 day data for all counties

## 5.0 Discussion

The main results from the different data sources and models are quite revealing. First it is clear that for the early unsmoothed data the large increase in risk in late March-early April favored a large asymptomatic rate and time dependence. This is likely due to the large variation over time from zero counts to high incidence. For the later period from April 2$^{nd}$ to June 29$^{th}$ an asymptomatic rate of 25% yielded a better fit and a model with both the poverty covariate and spatial residual CAR effect were best for either the crude data or the smoothed data. In general the smoothed data support a lower asymptomatic rate of 25%, inclusion of the county level poverty covariate and spatial structure, either as a spatially-structured intercept (model 5) or as a residual confounding (model 3). This further suggests that in the earlier period the time dependence in the risk in the unsmoothed data was heterogeneous with respect to county differences and so for the smoothed data spatial differentiation of risk was more obvious.

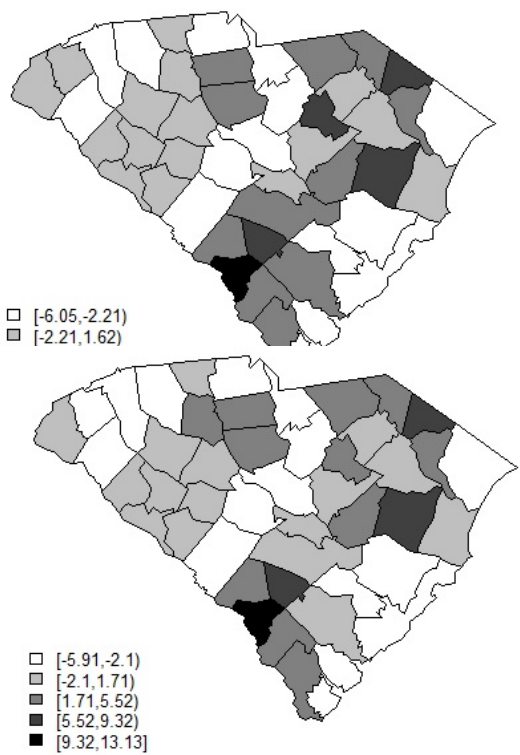

Figure 13 Posterior mean ICAR effect for model 3 for the smoothed 3 day average data April 2nd - June 29$^{th}$

Figure 13 displays the residual spatial ICAR effect for model 3 for the smoothed April-June data set. It is clear that the counties with greatest residual risk unexplained are rural in nature: the urban areas have much lower and negative residual effect. Figure 14 displays the spatially structured ICAR intercept effect for model 5 for the Jan-April smoothed data. This suggests that transmission is differentiated spatially and is higher in rural areas.

Figure 14 Posterior mean ICAR effect for the intercept in model 5 for the January -April smoothed data.

It is also notable that the neighborhood transmission model (model 2) was not selected for any of these datasets. This may suggest that at county level the resolution level does not support cross-boundary transmission. Finally it is also clear that % under the poverty line as a covariate is significant in these models so that regions of higher % poverty contribute more to the risk variation.

**5.1 Further Modeling considerations and limitations**

Currently available data are limited by a number of factors. Under-ascertainment of positive cases is a major concern. Asymptomatic or pre-symptomatic people are not counted. Further some patients may not attend for Covid-19 screening even if they are in fact positive for the disease. In addition the under-ascertainment could be spatially and temporally differentiated in that it could vary over time and between areas. For normal influenza seasons, health departments often use a scale factor to allow for the iceberg effect of not observing the true case load, when only lab confirmed cases are presented. This could be added for Covid-19 of course. As an alternative the true case load could be included within a Bayesian model as a latent effect.

An additional consideration is whether to model the death counts. If the focus is on the spread of Covid-19 then the death count is only a secondary outcome. In fact it is dependent on the true case count as opposed to infection process. A Poisson or binomial model could be assumed for this purpose. However as the focus here was on infection spread and case counts it has been assumed that death count is given and assumed fixed.

Another aspect of the data quality which has not been addressed here is the fact that the degree of testing carried out affects the number of positive cases found.

Assume the following, conditioning on the number of tests:

$y_{ij}^{sym} \sim bin(p_{ij}^p, n_{ij}^t)$

$p_{ij}^p$ : proportion of +ve tests and $n_{ij}^t$ is the total number of tests

The test positive proportion could be modeled via a logistic link such as: $\operatorname{logit}(p_{ij}^p)$ with dependence on a sum of lagged case numbers, and for the *i* th region the neighboring county sum of case numbers. The asymptomatic case numbers could be included as an additional effect. However, data for the number of tests is not currently publicly accessible for county level in SC and so this cannot be implemented at the moment, so this must remain a future development.

**6.0 Conclusions**

Some conclusion can be drawn from the analysis described here. The most important conclusions in this analysis of county level provincial data are as follows.

1) Smoothed data (at least 3 day averaged data) can yield different results from crude unsmoothed count data, especially with respect to time dependence and spatial smoothing. Smoothed data could induce greater spatial correlation and this was highlighted above.
2) The asymptomatic rate was varied in these models and was found to best fit when it was high (50%) for early unsmoothed data with a large spike in risk. More commonly 25% rate was favored for later data and the smoothed 3 day average data in general. CDC2020 lists a range of rates but cites 40% as the current best estimate
3) Poverty does provide better explanation of risk variation in these county level examples, and hence its surrogates of unemployment, low median income, and ethnicity differences could also be assumed to play a role. These additional predictors have not been included here but may help to reduce the residual spatial structure in the risk variation.
4) The value of spatio-temporal Bayesian modeling lies in the ability to flexible add both observed explanatory variables and unobserved latent effects to model disease risk. In our models we have achieved a reasonable measure of fit and made comparisons of different models with different assumptions, provided insight into asymptomatic rate and spatial differentiation of risk. The role of poverty is also clear. The ability to add cross-boundary transmission as a gravity effect is also a feature of the modeling, although in this case this was not selected in the best model.

**Acknowledgement**

We are indebted to Dr Diba Khan, Centers for Disease Control and Prevention (CDC), Office of Research and Methodology, for provision of data management support during a difficult time.

**Appendix**
Nimble code for the model 4
--------------------------------------------------------------------------------
```
COVModel4<-nimbleCode({
for (i in 1:M){
remc[i,1]<-0
remD[i,1]<-0
susc[i,1]<-susint[i]
muc[i,1]<-0.001*susc[i,1]
sym[i,1]~dpois(muc[i,1])
asym[i,1]<-0.33*sym[i,1]

LdevC[i,1]<--2*(sym[i,1]*log(muc[i,1]+0.001)-(muc[i,1]+0.001)-lfactorial(sym[i,1]))
}
for (i in 1:M){
for (j in 2: T){
remc[i,j]<-betaRc*sym[i,j]
remD[i,j]<-deaths[i,j]
susc[i,j]<-susc[i,j-1]-sym[i,j-1]-asym[i,j-1]-remc[i,j-1]-remD[i,j-1]
sym[i,j]~dpois(muc[i,j])
asym[i,j]<-0.33*sym[i,j]
log(muc[i,j])<-bet0[j]+log(susc[i,j]+0.001)+bet1*(log(sym[i,j-1]+0.001)+log(asym[i,j-1]+0.001))+bet2*percP[i]+b1[i]
LdevC[i,j]<--2*(sym[i,j]*log(muc[i,j]+0.001)-(muc[i,j]+0.001)-lfactorial(sym[i,j]))
}

muct1[i]<-muc[i,1]
muct2[i]<-muc[i,2]
muct3[i]<-muc[i,3]
muct4[i]<-muc[i,4]
muct5[i]<-muc[i,5]
muct6[i]<-muc[i,6]
muct7[i]<-muc[i,7]
muct8[i]<-muc[i,8]
muct9[i]<-muc[i,9]
muct10[i]<-muc[i,10]
muct40[i]<-muc[i,40]
muct46[i]<-muc[i,46]
muct50[i]<-muc[i,50]
muct60[i]<-muc[i,60]
muct70[i]<-muc[i,70]
remt1[i]<-remc[i,1]+remD[i,1]
remt2[i]<-remc[i,2]+remD[i,2]
remt3[i]<-remc[i,3]+remD[i,3]
remt4[i]<-remc[i,4]+remD[i,4]
remt5[i]<-remc[i,5]+remD[i,5]
remt6[i]<-remc[i,6]+remD[i,6]
remt7[i]<-remc[i,7]+remD[i,7]
remt8[i]<-remc[i,8]+remD[i,8]
remt9[i]<-remc[i,9]+remD[i,9]
remt10[i]<-remc[i,10]+remD[i,10]
```

```
sust1[i]<-susc[i,1]
sust2[i]<-susc[i,2]
sust3[i]<-susc[i,3]
sust4[i]<-susc[i,4]
sust5[i]<-susc[i,5]
sust6[i]<-susc[i,6]
sust7[i]<-susc[i,7]
sust8[i]<-susc[i,8]
sust9[i]<-susc[i,9]
sust10[i]<-susc[i,10]
}

for (j in 1:T){
mucrich[j]<-muc[40,j]
mucchar[j]<-muc[10,j]
muchor[j]<-muc[26,j]
mucbea[j]<-muc[7,j]
mucker[j]<-muc[28,j]
mucand[j]<-muc[4,j]
mucsum[j]<-muc[43,j]
mucflor[j]<-muc[21,j]
}
b1[1:46] ~ dcar_normal(adj[1:L], wei[1:L], num[1:M], tau.b1,zero_mean=1)
  for(k in 1:L) {wei[k] <- 1 }

DevC<-sum(LdevC[1:M,1:T])
for (k in 1:T){
bet0[k]~dnorm(0,tau0)}
tau0~dgamma(2,0.5)
bet1~dnorm(0,tau1)
tau1~dgamma(2,0.5)
tau.b1~dgamma(0.01,0.01)
betaRc<-0.1
bet2~dnorm(0,tau2)
tau2~dgamma(2,0.5)
#R0<-exp(bet0)/betaRc
}
```